\documentstyle[aps]{revtex}
\begin{document}

\draft

\title{Self-gravitating domain walls and the thin-wall limit }

\author{Rommel Guerrero\footnote{rommel@uicm.ucla.edu.ve}}
\address{\it Departamento de Matem\'aticas, 
Universidad Centroccidental Lisandro Alvarado, Barquisimeto, Venezuela}
\author{        Alejandra Melfo\footnote{melfo@ula.ve} and
        Nelson Pantoja\footnote{pantoja@ula.ve}}
\address{ {\it Centro de Astrof\'{\i}sica Te\'orica, Universidad de 
Los Andes, M\'erida, Venezuela }}
\maketitle

\begin{abstract}

We analyse the distributional thin wall limit of self gravitating
scalar field configurations representing 
thick domain wall geometries. We show that thick wall
solutions can be generated by appropiate scaling of the thin
wall ones, and obtain an exact solution
for a domain wall that interpolates between $AdS_4$ asymptotic
vacua and  has a well-defined thin wall limit. 
Solutions representing scalar field configurations
obtained via the same scaling but that do not have  a thin wall limit 
are also presented. 
\end{abstract}

\pacs{
04.20.-q, 
11.27.+d  
}

\vspace{0.3cm}

\section{Introduction}

The gravitational properties of domain walls have been 
studied in the past due to their striking implications for cosmology. 
Recently, however, they have been object of intense investigation for
different reasons. One the one hand, it has been pointed out that
four-dimensional gravity can be realized on a thin wall interpolating
between $AdS$ spacetimes \cite{rs99}.  In addition, 
wall configurations are
relevant for the study of RGE flows in the context of $AdS$/CFT
correspondence \cite{m98}. 

The first attempts to study  these gravitational properties were based
in the so-called thin wall limit \cite{v83,is84}. 
In a four-dimensional spacetime, the
wall is treated as an infinitely thin (2+1) surface. The spacetime outside
the wall is given by vacuum solutions to the Einstein field equations
with a planar symmetry, and one can use the Darmois-Israel
\cite{israel} thin wall formalism
to match solutions across the surface.  Under this approximation, it is
possible to find exact solutions representing an infinitely thin wall
with an associated surface energy density. Spacetimes containing a
thin domain wall have therefore distributional curvatures and 
energy-momentum tensors, proportional to delta functions supported 
in the wall's surface. 

However, as pointed out in \cite{bcg99}, these thin
walls may be very artificial constructions in the sense that they 
do not necessarily correspond to the appropriate limit of a
 thick domain wall. Thick domain walls are solutions to the coupled
Einstein-scalar field equations interpolating between minima of a
potential with a spontaneously broken discrete symmetry. In order for
this thick walls to have a thin wall limit an appropriate
 distributional treatment of the curvature and
energy-momentum tensors is required. As is well known, computing the
curvature tensor from a metric requires nonlinear operations which are
not defined within the framework of distribution theory, and this
imposes strong constraints in the class of metrics whose curvature
tensors make sense as distributions \cite{gt87}. 

The first exact thick wall solution in D=4 was obtained by Goetz
\cite{g90}. The only other solution in the literature is the one of
Ref.  \cite{m93,gm99}, for a thick wall without reflection symmetry. 
Numerical solutions were found in \cite{bcg99}
(see also references therein).  The thin wall limit of these solutions
has not been studied. All of these solutions represent walls
in spacetimes without a cosmological constant.

The purpose of this paper is twofold. First, we analyze the thin wall
limit of thick wall solutions. We find that the solution of
\cite{g90} has a well-defined thin wall limit. We also show that
this solution can be obtained by an appropriate scaling of a solution
of Einstein's equations in vacuum with a planar symmetry. Using this
fact, we generate a number of exact solutions to the Einstein-scalar
field equations using the vacuum solutions. We then show that the only
wall that can be considered a thick domain wall and that possesses  a
thin wall limit is the one of Ref.  \cite{g90}.

Secondly, we look for solutions representing a domain wall
embedded in a four-dimensional spacetime with a negative 
cosmological constant. We show how the scaling procedure
can be modified to generate such solutions, and find an exact solution
for a thick domain
wall interpolating between two $AdS_4$ vacua. This wall is then shown
to have a curvature and energy-momentum tensor well defined as
distributions and the corresponding distributional thin-wall limit is
obtained. The  possibility of obtaining this solution within the supergravity 
inspired first order formalism of Ref. \cite{st99,dfgk00} is
 also investigated, and  
the scalar field potential is shown to satisfy the requirements for
 the existence of stable $AdS$ vacua.

 The paper is organized as follows. In the next section we study the
thin wall limit of thick wall solutions. In Section III, we show how
exact solutions can be found by scaling thin wall spacetimes, and show
that the new solutions found do not represent true domain walls. In
Section IV, a new solution for a domain wall in an $AdS$ spacetime is
found and analyzed. We summarize our results on Section V, and include
an Appendix for the reader interested in the details of the
distributional analysis.

\section{From thick to thin domain walls}

Consider the spacetime ($R^4$, {\bf g}), where the metric in a
particular coordinate system is given by

\begin{equation}
g_{ab} = \cosh(\beta x/\delta)^{-2 \delta} \left[ -dt_a dt_b + dx_a dx_b
+ e^{2 \beta t}( dy_a dy_b + dz_a dz_b)\right]
\label{metric}
\end{equation}
where $\beta$ and $\delta$ are constants. In Ref. \cite{g90,m93,gm99},
 this spacetime has been shown to be the one generated by 
a ``thick'' domain wall, i.e. it is a solution to the coupled
 Einstein-scalar field equations:
\begin{equation}
R_{ab} - \frac{1}{2} g_{ab} R = 8 \pi T_{ab}\quad ,
\label{e.f.e.}
\end{equation}
\begin{equation}
T_{ab} = \nabla_a \phi \nabla_b \phi - g_{ab}\left(\frac{1}{2}
\nabla_c \phi \nabla^c \phi + V(\phi)\right)
\label{tmunu}
\end{equation}
and
\begin{equation}
\nabla_a\nabla^a\phi - \frac{dV(\phi)}{d\phi} = 0 \quad ,
\label{k.g.}
\end{equation}
with
\begin{equation}
\phi = \phi_0\tan^{-1}\left(\sinh(\beta x/\delta)\right)\, ,
\quad  \phi_0 \equiv \sqrt{\frac{\delta (1 - \delta)}{4 \pi}}\label{phi1}
\end{equation}
and
\begin{equation}
V(\phi) = \frac{1 + 2 \delta}{8 \pi \delta} \beta^2 \left[\cos\left(\phi/\phi_0\right)
\right]^{2(1-\delta)}
\label{ve1}
\end{equation}
where $0<\delta<1$. This solutions represent a domain wall of (finite)
thickness $\delta$. The scalar field takes values $\pm \phi_0 \pi/2$ at
$x=\pm\infty$, corresponding to   two consecutive minima of the
potential, and interpolates smoothly between
these values at the origin.

 The metric (\ref{metric}) is only one of many possible thick
domain wall solutions which can be obtained under the requirements: i)
$V(\phi)\geq 0$, ii) $g^{ab} \partial_a\phi\partial_b\phi > 0$ and iii)
reflection symmetry. However, (\ref{metric},\ref{phi1},\ref{ve1}) is the only
analytic solution encountered in the literature so far \cite{g90} (for
a study of its properties, see \cite{gm99,wl94}).  Numerical
treatments can be found in \cite{bcg99} and references therein.

Next,  consider the $\delta\to 1 $ and $\delta \to 0$ limits of this spacetime:

\vspace{0.5cm}
\noindent {\bf 1.}  For $\delta\to 1$  we have $\phi = 0,
V(\phi) = 3 \beta^2$, and the metric

\begin{equation}
g_{ab} = \left( \cosh(\beta x)  
\right)^{-2}\left[ -dt_a dt_b + dx_a dx_b + e^{2 \beta t}(
dy_a dy_b + dz_a dz_b)\right]
\label{ma1}
\end{equation}
turns out to be a solution to the Einstein field equations
for the vacuum with a cosmological constant term

\begin{equation}
R_{ab} - \frac{1}{2} g_{ab} R +  g_{ab} \Lambda = 0 \quad ,
\label{e.f.e.lambda}
\end{equation}
where $\Lambda = V(\phi)|_{\delta=1} = 3 \beta^2 $. Under the assumption that
the energy-momentum tensor of a thin  wall can be approximated by a
cosmological constant outside the wall (where the nearly-constant
potential term is dominating), the solution (\ref{ma1}) of
(\ref{e.f.e.lambda}) can be interpreted  as representing the spacetime
at some distance from the wall \cite{cgs93}. However, notice that there is no
thin wall  at the origin or elsewhere, the metric being well-defined in
all spacetime and having a non-singular curvature tensor. In fact,
the limit $\delta\to 1$ can be considered as
representing a wall of infinite thickness.

\vspace{0.2cm}
\noindent {\bf 2.}  We turn now to the $\delta\to 0$
limit. In order to compute this limit we  consider the curvature
tensor and the Einstein tensor as distributional tensor fields. As
is well known, the curvature tensor being non linear does not
make sense in general as a distribution. However, the metric
(\ref{metric}) is a smooth metric that belongs to the class of
regular metrics \cite{gt87}, and for a regular metric the curvature
tensor makes sense as a distribution. Since any contraction of a
distribution is also a distribution, for a regular metric the
Einstein tensor is well defined as a distribution.

Taking the $\delta\to 0$ limit, we find

\begin{equation} \lim_{\delta\to 0} g_{ab} = e^{-2 \beta |x|} (
-dt_adt_b+ dx_adx_b 
+ e^{2 \beta t}(dy_ady_b + dz_adz_b)), \label{vacio} \end{equation}
which for $x<0$ and $x>0$ is a vacuum solution of the Einstein
field equations \cite{cgs93}, and

\begin{equation} \lim_{\delta\to 0} G^a_{\, b} = - 4 \beta \delta(x)\left[
\partial t^a dt_b + \partial y^a dy_b + \partial z^a dz_b \right]
 \end{equation}

This means that the spacetime ($R^4$, g), with the metric
given by (\ref{metric}), can be identified in the limit $\delta\to 0$
with the spacetime $(R^4,g)$, with $g$ given by (\ref{vacio}),
generated by a thin domain wall with energy-momentum tensor given
by

\begin{equation} 8 \pi T^a_b =  - 4 \beta \delta(x)\left[
\partial t^a dt_b + \partial y^a dy_b + \partial z^a dz_b \right]
 \label{tmunudelta} \end{equation}

As expected, (\ref{tmunudelta}) can be obtained from (\ref{vacio})
by using the formalism of Israel \cite{israel} to treat surface
layers.

Actually, to be rigorous, one should prove that  the metric
(\ref{metric}) provides a sequence of metrics that satisfies the
required convergence condition of \cite{gt87}. Then the limit of
the curvature tensor exists and is the curvature tensor of the
limit metric. We leave this rather technical proof for the
Appendix.

Remarkably enough, the metrics (\ref{metric}), solution to the
coupled Einstein-scalar field equations,  and (\ref{ma1}), vacuum
solution, can be rewritten simply as

\begin{eqnarray}
{}_\delta g_{ab} &=& f^{2 \delta}(x/\delta) [ -dt_a dt_b + dx_a dx_b +
e^{ 2 \beta t}(dy_a dy_b + dz_a dz_b) ]  \\
&  & \nonumber \\
g_{ab} &=& f^{2}(x) [ -dt_a dt_b + dx_a dx_b + e^{ 2 \beta t}(dy_a
dy_b + dz_a dz_b) ]
\end{eqnarray}
respectively, where $f(x) = \cosh(\beta x)^{-1}$. This opens up the interesting
possibility of generating ``thick wall'' solutions from the ``thin
wall'' ones, i.e. by scaling vacuum solutions with positive, negative or null
cosmological constant. In the next section, we will explore this
possibility.

\section{From thin to thick walls}

We wish to find solutions to the coupled Einstein-scalar field equations
(\ref{e.f.e.}-\ref{k.g.}) with a planar symmetry, where the scalar
field $\phi$ is static. In its most general
form, the metric can be written

\begin{equation}
g_{ab} = f(x)^2 [-dt_a dt_b + dx_a dx_b] + B(x,t)^2 [(1 - \kappa r^2)^{-1} dr^2 + r^2 d\varphi^2]
\label{metricgeneral}
\end{equation}
where $\kappa$ is the curvature of the $x,t=const.$ surfaces.
Requiring that $\phi$ be a function of the coordinate perpendicular to
the wall only, and that the spacetime be boost-invariant in directions
parallel to the wall, it can be shown \cite{cgs93} that

\begin{equation}
B(x,t) = f(x)  C(t)
\label{be}
\end{equation}
where the function $C(t)$ satisfies (dot indicates differentiation
with respect to $t$)
\begin{equation}
\frac{\ddot{C}(t)}{C(t)} = \frac{\dot{C}(t)^2}{C(t)^2}  +
\frac{\kappa}{C(t)^2} = \beta^2
\end{equation}

The time-dependence of the solutions depends on the sign of the constant
$\beta^2$. For positively curved ($\kappa>0$) or planar  ($\kappa=0$) walls,
$\beta^2$ is
positive and we have
\begin{eqnarray}
C(t) &=&  e^{\beta t}, \quad {\rm when}\, \kappa =0 \label{tk0} \\
& & \nonumber \\
C(t) &=& \cosh(\beta t), \; \beta^2 = \kappa , \quad  {\rm when}\, \kappa >0\label{tkpos}
\end{eqnarray}
whereas for negatively curved ($\kappa< 0$) walls the sign of $\beta^2$ is
arbitrary, and three solutions are possible

\begin{eqnarray}
C(t) &=  \sinh(\beta t),&\quad  \beta^2 = - \kappa  \label{tkneg1}\\
 &= \sqrt{-\kappa} t, &\quad   \beta^2 = 0 \label{tkneg2}\\
&=\sin(\beta t),&\quad \beta^2 = \kappa \label{tkneg3}
 \end{eqnarray}

Then the system  (\ref{e.f.e.}-\ref{k.g.}) reduces to just two
equations for the potential and the scalar field. Define (prime denotes
derivative respect to $x$)
\begin{equation}
u(x) \equiv \frac{f(x)'}{f(x)} \, ;
\end{equation}
then
\begin{eqnarray}
\phi'^2 &=& \frac{1}{4\pi} \left[ - u(x)' + u(x)^2 - \beta^2 \right] \label{phiprime} \\
& & \nonumber \\
V(\phi)&=&\frac{1}{8\pi f^2}\left[- u(x)' -
2 ( u(x)^2 + \beta^2) \right]  \label{potential} \, .
\end{eqnarray}

The solution of Refs. \cite{g90,gm99} is found with  $f(x)=
[\cosh(\beta x/\delta)]^{-\delta}$, with $\kappa=0$ and $C(t)$ given by (\ref{tk0}), while as noted in the previous section,
 a vacuum solution with  a cosmological constant is obtained with
$f(x)=[\cosh(\beta x)]^{-1}$ and the same curvature and time-dependence.

 We have found that this is a general result: {\em the system
(\ref{phiprime}-\ref{potential}) can be integrated with the scaled function}
\begin{equation} f(x)= f_0(x/\delta)^\delta ,
\label{efe0}\end{equation}
{\it where $f_0(x)$ is a solution to the Einstein
field equations in vacuum with a cosmological constant for the metric
(\ref{metricgeneral})}.

This is easily shown. Substituting(\ref{efe0}) in (\ref{phiprime})

\begin{equation}
\phi'^2 =  \frac{1}{4\pi} \left[ - \frac{1}{\delta} u_0(x/\delta)' +
 u_0(x/\delta)^2 - \beta^2 \right]
\label{phiprime0}
\end{equation}
where now prime denotes derivative respect to the argument. Since $f_0$ satisfies the
Einstein equations (\ref{e.f.e.lambda}), we have
\begin{eqnarray}
u_0(x/\delta)' + 2 (u_0(x/\delta)^2 - \beta^2 ) &=& - \Lambda f_0(x/\delta)^2\\
 u_0(x/\delta)^2 - \beta^2 - u_0(x/\delta)' = 0 \; ,
\end{eqnarray}
and substituting in (\ref{phiprime0},\ref{potential})
\begin{equation}
\phi = \sqrt{\frac{\delta(1-\delta)}{4\pi} \frac{\Lambda}{3}}\int_{x_0}^{x/\delta}
f_0(\xi) d\xi \; ,
\end{equation}
\begin{equation}
V(x)= \frac{1}{8\pi} \frac{\Lambda}{3}\left(\frac{1 + 2
\delta}{\delta}\right)f_0(x/\delta)^{2(1-\delta)} \; .
\end{equation}

It is then possible to generate solutions representing a
self-gravitating scalar field wall using the vacuum solutions of
Ref. \cite{cgs93}. The time-dependence of the metric and the curvature
 of the wall  will be preserved.

For vacuum solutions with (\ref{metricgeneral},\ref{be}) and  negative
cosmological constant we get: 

\begin{enumerate}

\item with
$f_0(x) = [\sinh(\beta x)]^{-1}$, corresponding to a vacuum solution
with $\Lambda = -3 \beta^2$,  a solution of (\ref{e.f.e.}-\ref{ve1}) with
\begin{eqnarray}
f(x)&=& [\sinh(\beta x/\delta)]^{-\delta} \nonumber \\
\phi(x) &=& \phi_0 \coth^{-1}[\cosh(\beta x/\delta)], \quad \phi_0\equiv
-\sqrt{\frac{\delta(\delta-1)}{4 \pi }} \nonumber \\
V(\phi) &=& \frac{\beta^2}{8 \pi}\frac{2 \delta + 1}{ \delta}
[\sinh(\phi/\phi_0)]^{2(1-\delta)} 
\end{eqnarray}

As in the vacuum case, $C(t)$ is given by either (\ref{tk0}),
 (\ref{tkpos}), or (\ref{tkneg1}) when the
curvature of the wall is zero, positive or negative respectively.

\item with
$f_0(x) = [1 + \alpha x]^{-1}$, corresponding to a vacuum solution
with $\Lambda = -3 \alpha^2$, a solution of (\ref{e.f.e.}-\ref{ve1}) with
\begin{eqnarray}
f(x)&=& [1 + \alpha x/\delta]^{-\delta} \nonumber \\
\phi(x) &=& \phi_0 \ln(1+ \alpha x/\delta), \quad \phi_0\equiv
\sqrt{\frac{\delta(\delta-1)}{4 \pi }} \nonumber \\
V(\phi) &=& \frac{\alpha^2}{8\pi}\frac{2 \delta + 1}{ \delta}
exp[2(\delta-1)\phi/\phi_0]
\end{eqnarray}

In this case the plane
wall corresponds to a static solution $C(t)=1$,  but a non-static wall
is possible with negative curvature and  $C(t)$  given by (\ref{tkneg2}),

\item with (\ref{metricgeneral},\ref{be}) and
$f_0(x) = [\cos(\beta x)]^{-1}$,  vacuum solution with $\Lambda =  -3
\beta^2$, a solution of (\ref{e.f.e.}-\ref{ve1}) with
\begin{eqnarray}
f(x)&=& [\cos(\beta x/\delta)]^{-\delta} \nonumber \\
\phi(x) &=& \phi_0 \tanh^{-1}[\sin(\beta x/\delta)], \quad \phi_0\equiv
\sqrt{\frac{\delta(\delta + 1)}{4 \pi }} \nonumber \\
V(\phi) &=& \frac{\beta^2}{8 \pi} \frac{2 \delta -1}{\delta}
[\cosh(\phi/\phi_0)]^{2(1-\delta)}
\end{eqnarray}
\end{enumerate}

The wall must have in this case negative curvature and $C(t)$ given by
(\ref{tkneg3}).

 A fourth non-trivial
vacuum solution exists with $\Lambda=0$, namely
\begin{equation}
f(x) = e^{\pm \beta x}
\end{equation}
but obviously it cannot generate a thick wall solution by the same
scaling procedure.

All of these solutions have an energy-momentum tensor of the form

\begin{equation}
 T^a_{\,b} = -\rho(x) [\partial t^adt_b+ \partial y^ady_b + \partial
z^adz_b] + p(x)\partial x^adx_b
\end{equation}
compatible with a static scalar field wall. In the first solution, the
parameter $\delta$ can  be interpreted as the wall's thickness, just as in
the solution of the previous section. However, notice that 
this spacetime contains a singularity which seems to be much worse
than the one produced by a source concentrated on a thin wall. 
In this example the metric is not regular, and we cannot assign to it a
distributional source following the approach of \cite{gt87}.

In fact, none of these solutions represents a {\em domain}
wall. Namely, none of the potentials above has minima or is even
bounded from below. These wall solutions are not topologically
protected, and their stability is thus not guaranteed.
For example, take
case 1 above. Although $\phi$ takes
constant values at infinity, it does not interpolate smoothly between
them.

 Notice also that far
from the walls one recovers the vacuum solutions. Again, take case 1:
the metric has the same asymptotic behavior as
the domain wall solution of Section II,
\begin{equation}
g_{ab} \to  
 e^{-\beta |x|} [ -dt_a dt_b + dx_a dx_b + e^{ 2 \beta
t}(dy_a dy_b + dz_a dz_b) ] \quad {\rm when}\; {\beta |x| \to \infty} .
\end{equation}
 This is an important point: finding a vacuum solution to the Einstein
equations with planar symmetry, and then using the thin shell
formalism of Israel may produce a thin wall which is not the thin limit
of a scalar thick domain wall. In this sense, the
thin wall solution is less artificial if it can be obtained as the
limit of a smooth configuration. 

It is possible however to find other solutions to
(\ref{phiprime},\ref{potential}), not generated by thin wall ones,
 that do represent a thick  domain wall with a well defined 
thin wall limit, as we do in the next section.

\section{A thick domain wall with cosmological constant}

In this section we consider thick domain walls embedded in a spacetime
with negative cosmological constant $\Lambda$. The case $\Lambda < 0$
is particularly interesting because the positive effective
gravitational mass density of $AdS_4$ spacetime may counteract the
negative effective gravitational mass density of the domain wall.  On
the other hand, a certain type of these solutions may be realized as
supersymmetric bosonic field configuration \cite{cgr92}. Clearly,
for $\Lambda<0$ we are looking for domain wall solutions where the
effective potential $V_{\rm eff} \equiv V + \Lambda/8\pi$ is not
necessarily positive-definite, requiring only that it is bounded from below.

Assuming a conformally flat symmetric {\em ansatz} for  the metric
\begin{equation}
g_{ab} = f(x)^2 [-dt_a dt_b + dx_a dx_b + dy_a dy_b +dz_a dz_b]
\end{equation}
and also that the scalar field depends only on $x$, the equations of
motion are (\ref{phiprime}, \ref{potential}) where $\beta =0$, and with
the addition of a cosmological constant term
$-\Lambda/8\pi$ to the right-hand-side of  (\ref{potential}).

 We  look for reflection symmetric domain wall solutions for
$\Lambda<0$. Under the requirements
\begin{description}
\item[{\it i)}] $V(\phi) > 0$
\item[{\it ii)}] $\lim_{|x|\to 0}|\phi(x)'| = 0$
\item[{\it iii)}]$\phi'(0) = 0$ (nonsingular solution)
\item[{\it iv)}]$(f^2)'|_{x=0} = 0$ (reflection symmetry)
\end{description}
we find the solution
\begin{eqnarray}
f(x) &=& (1 + \alpha^2 x^2)^{(-1/2)} \; , \\
\phi(x) &=& \phi_0 \tan^{-1}{(\alpha x)} \; ,\\
V(\phi) &=&\frac{1}{2\pi}  \alpha^2 \cos^2{ (\phi/\phi_0)} \; ,\\
\Lambda &=& - 3 \alpha^2\; .
\end{eqnarray}
where $\phi_0 = \sqrt{1/4\pi}$. This solution represents a class of
gravitating domain walls that interpolate smoothly  between two minima
of the potential $V(\phi)$, the spacetime being asymptotically
$AdS_4$. With the coordinate change
\begin{equation}
\alpha\xi = \sinh^{-1}{(\alpha x)} \, ,
\end{equation}
the line element takes the form
\begin{equation}
g_{ab} = \cosh^{-2}{(\alpha \xi)} [-dt_a dt_b  + dy_a dy_b
+dz_a dz_b] + d\xi^2 \, ,
\end{equation}
which asymptotically behaves as $ AdS_4$
\begin{equation}
g_{ab} \to  4 e^{-2 \alpha |\xi|}
 [-dt_a dt_b  + dy_a dy_b +dz_a dz_b] + d\xi^2 \quad {\rm
when}\; {\alpha |\xi| \to \infty}  . 
\end{equation}

We now wish to consider the thin wall limit. However, $AdS_4$
domain wall solutions generically have two free parameters: one
for the asymptotic $AdS$ curvature and one for the wall's width.
Thus, in order to introduce a second parameter in the solution
found, we make a scaling of the metric tensor as in the previous
section.

Consider the scaled metric
\begin{equation}
g_{ab}dx^a dx^b = \cosh^{-2 \delta}{(\alpha \xi/\delta)} [-dt_a dt_b  + dy_a dy_b +dz_a dz_b] +
d\xi^2
\label{scaledm}
\end{equation}
We find this time
\begin{eqnarray}
\phi(\xi) &=& \phi_0 \tan^{-1}{\sinh(\alpha \xi/\delta)} \; , \label{scaled1}\\
V(\phi) &=& \frac{1}{8\pi}\alpha^2\left(3 + \frac{1}{\delta}\right) \cos^2{
(\phi/\phi_0)} \; ,\label{scaled2}
 \\
\Lambda &=& - 3 \alpha^2\;  \label{scaled3} \end{eqnarray}
with $\phi_0 = \sqrt{\delta/4\pi}$. It is easy to see that the metric
(\ref{scaledm}) is a regular metric in the differentiable
structure provided by the coordinate chart $\{t,\xi,y,z\}$. It
follows that the curvature and Einstein tensor fields are well
defined as distributions.

Computations analogous to the ones in the Appendix show
that
\begin{equation} \lim_{\delta\to 0}g_{ab} = 4 e^{-2\alpha|\xi|}
 \left( -dt_a dt_b + dy_a dy_b + dz_a dz_b\right) 
 + d\xi_a d\xi_b, \end{equation}
which is also a regular metric and

\begin{equation} \lim_{\delta\to 0}(G^a_{\, b} + \Lambda g^a_{\, b}) =
- 4\alpha \delta(\xi)  \left( \partial_t^a dt_b + \partial_y^a dy_b +
\partial_z^a dz_b\right) \end{equation}
where $\xi=0$ is the codimension one hypersurface where the thin
wall is located and $\Lambda$ is given by (\ref{scaled3}).

Thus, we have found a two-parameter family of self-gravitating
scalar fields with a thick domain wall profile interpolating
between two $AdS_4$ vacua.  Furthermore, these solutions have a
distributional curvature tensor with a well-defined thin wall
limit in the sense of Ref. \cite{gt87}.

As stated above, the domain wall spacetime considered in this
section is asymptotically $AdS_4$. It is known that restrictions
on the potential are to be imposed from the requirement that there
exist a stable $AdS$ vacuum \cite{b84,t84}. In $D$ dimensions, for
a model with Lagrangian density 
\begin{equation} {\cal L} = \sqrt{-g} \left[
\frac{1}{2} R -  \frac{1}{2}(\partial \phi)^2 - {\cal V}(\phi)
\right] \end{equation} 
vacuum stability requires ${\cal V}$ to take the form
\begin{equation} {\cal V} = 2(D-2)\left[ (D-2) \left(\frac{dW}{d\phi}\right)^2
- (D-1)W^2 \right] \end{equation}
 where $W(\phi)$ is any function with at
least one critical point \cite{t84}. Critical points of $W$ are
also critical points of ${\cal V}$ and in the context of
supergravity theories the critical points of $W$ yield stable
$AdS$ vacua \cite{st99}.

For the domain wall solution (\ref{scaledm}-\ref{scaled3}), this
is equivalent to requiring that

\begin{equation} {\cal V} = V_{\rm eff} = V + \Lambda = 4 \left[ 2
\left(\frac{dW}{d\phi}\right)^2 - 3 W^2 \right] \end{equation} 
with $V$ and
$\Lambda$ given by (\ref{scaled2}) and (\ref{scaled3})
respectively. It follows that in this case 
\begin{equation} W(\phi) =
\frac{1}{2}\beta \sin(\phi/\phi_0) \label{superpot} \end{equation} 
whose
critical points are $\phi = \pm \pi \phi_0/2$. Whether a
supergravity theory can be constructed so that the supersymmetry
conditions lead to (\ref{superpot}) is a question beyond the scope
of this paper. But since the critical points of  (\ref{superpot})
are the asymptotic values of $\phi$, as given by (\ref{scaled1}),
this suggests that these asymptotic $AdS$ vacua are stable.

It should be noted that 
 (\ref{scaledm}-\ref{scaled3}) can be parametrized by
(\ref{superpot}), so we could have found it using the first order
formalism of \cite{st99,dfgk00}. However, (\ref{scaled1}) is not
the familiar kink which one usually encounters in the literature,
whether supersymmetric or not \cite{bcy01,dfgk00}. Finally, in Ref. \cite{g99}
an example with a superpotential similar to (\ref{superpot}) has been
considered in $D=5$, in the study of how four-dimensional gravity
arises on a thick wall interpolating between two $AdS_5$ spacetimes.

\section{Concluding remarks}

We have studied the thin-wall limit of  thick domain wall solutions in
a (3+1)-dimensional spacetime. We have shown that the 
solution (\ref{metric}) of Ref. \cite{g90}
represents a spacetime with a regular metric in the
sense of Ref \cite{gt87}, and that the thin wall limit can be taken
rigorously in distribution theory. Not surprisingly, in the thin-wall
limit solution (\ref{metric}) becomes the well-known thin wall solution 
of \cite{v83}.
We have also demonstrated that this thick solution can be obtained by
appropriately scaling thin, i.e. vacuum, ones. However, although other
solutions to the Einstein-scalar field coupled equations can be
systematically obtained by the same procedure, it does not follow
that these new solutions are thick walls, even if they have the
appropriate asymptotic behavior far from the origin  and the stress-energy
tensor of a wall. The scalar field potential is in general not bounded
from below and the scalar field configurations are not topologically
protected, thus probably unstable.  

Using a similar scaling procedure, we have obtained a solution
representing a thick  domain wall embedded in a $AdS_4$
spacetime. The cosmological constant, as expected, is related to the
wall' s surface energy density. This solution was shown to have a
thin-wall limit, with a stress-energy tensor which is well defined as a
distribution.  The potential is positive definite and the
scalar field smoothly interpolates between two $AdS$ vacua. 
Moreover, the scalar field potential for this solution has been shown
to  satisfy the requirements for the existence of stable $AdS$ vacua,
being derivable from a superpotential function. The possible connection
with supergravity theories, and the stability of the solution under
perturbations are currently under investigation.

\section{Acknowledgements}

We wish to thank H. Rago and A. Skirzewski for discussions.  
This work was financed by CDCHT-ULA under project C-1066-01-05-B.
 
\section{Appendix}

In this paper we use the definition of tensor distribution given
by Geroch and Traschen. The reader is referred to \cite{gt87} for
details.

{\bf Definition:} A symmetric tensor field $g_{ab}$ will be called a
regular metric provided that {\it i)} $g_{ab}$ and  $(g^{-1})^{ab}$
exist everywhere and are locally bounded and  {\it ii)} the weak
derivative of $g_{ab}$ in some smooth metric $\eta_{ab}$ exists and is
locally square integrable.

The curvature tensor and the Einstein tensor of a regular metric make
sense as distributions, therefore it makes sense to write Einstein's
equations with a distributional energy-momentum tensor. Furthermore,
these idealized matter sources are necessarily concentrated on
submanifolds of codimension of at most one.

First define conveniently the smooth tensor fields

\begin{eqnarray}
S_{ab} &\equiv&  -dt_a dt_b +
dx_a dx_b
+ e^{2 \beta t}( dy_a dy_b + dz_a dz_b)\\
(S^{-1})^{ab} &\equiv& -\partial_t^a\partial_t^ b +
\partial_x^a\partial_x^ b 
+ e^{-2 \beta t}(\partial_y^a\partial_y^ b +\partial_z^a\partial_z^ b 
 )
\end{eqnarray}
Now let
\begin{equation}
{}_n g_{ab}  = \cosh(\beta n x)^{-2/n} S_{ab}\, , \;\quad  g_{ab} 
 = e^{-2\beta |x|}S_{ab}
\label{Anmetric}
\end{equation}
and
\begin{equation}
 ({}_n g^{-1})^{ab} = \cosh(\beta n x)^{2/n} (S^{-1})^{ab}, \quad 
(g^{-1}) ^{ab}  = e^{2\beta |x|}(S^{-1})^{ab}
 \label{Ametrici}
\end{equation}

Let $U^{ab}$ be a test tensor field defined on $R^4$. We have 
\begin{equation} 
{}_n g_{ab} U^{ab} = \cosh(n \beta x)^{-2/n}\left[ -U^{tt} + U^{xx}
+ e^{2 \beta t} (U^{yy} + U^{zz})\right]\omega_\eta
\label{Adnmetric} \end{equation}
and
\begin{equation} 
g_{ab} U^{ab} =  e^{-2\beta |x|} \left[ -U^{tt} + U^{xx}
+ e^{2 \beta t} (U^{yy} + U^{zz})\right]\omega_\eta \, .
\label{Admetric} \end{equation}
Clearly,  ${}_n g_{ab} $ and $g_{ab}$ are locally bounded. Let  $U_{ab}$
be a test tensor field defined on $R^4$. We have
\begin{equation} 
({}_n g^{-1})^{ab} U_{ab} =  \cosh(n \beta x)^{2/n}\left[ -U_{tt} + U_{xx}
+ e^{-2 \beta t} (U_{yy} + U_{zz})\right]\omega_\eta
\label{Adnmetrici} \end{equation}
and
\begin{equation} 
(g^{-1})^{ab} U_{ab} =  e^{-2\beta |x|} \left[ -U_{tt} + U_{xx}
+ e^{-2 \beta t} (U_{yy} + U_{zz})\right]\omega_\eta
\label{Admetrici} \end{equation}
Hence  $({}_n g^{-1})^{ab} $ and $(g^{-1})_{ab}$ are locally bounded
also. 

Now choose as a smooth derivative operator $\nabla_a$, the one compatible with
the Minkowski metric $\eta_{ab}$ and let $U^{cab}$ be a test tensor field on
$R^4$. The weak derivative in $\eta_{ab}$ of   ${}_n g_{ab}$ and  $g_{ab}$
exist everywhere and are  given by

\begin{equation}
\nabla_c ({}_n g_{ab})[ U^{cab}] \equiv - {}_n g_{ab}[\nabla_c U^{cab}] =
\int_{R^4} {}_n W_{cab}U^{cab}\omega_\eta
\end{equation}
and the equivalent expression for  $g_{ab}$,
where
\begin{equation}
{}_n W_{cab} = 2 \beta \cosh(n \beta x)^{-2/n}\left\{\tanh(n \beta x) dx_c
[-dt_a dt_b  
dx_a dx_b + e^{2 \beta t}(dy_a dy_b + dz_a dz_b) ]
-  e^{2 \beta t} dt_c (dy_a
dy_b + dz_a dz_b)   \right\}
\label{Anw}
\end{equation}
and
\begin{equation}
W_{cab} =\left\{\begin{array}{ll}
2 \beta e^{2\beta x} [dx_c
(-dt_a dt_b + dx_a dx_b) + e^{2 \beta t} (dt_c + dx_c) (dy_a
dy_b + dz_a dz_b)],   & x < 0 \\ & \\
2 \beta e^{-2\beta x} [-dx_c
(-dt_a dt_b + dx_a dx_b) + e^{2 \beta t} (dt_c - dx_c) (dy_a
dy_b + dz_a dz_b)],   & x > 0 
\end{array} \right.
\label{Aw}
\end{equation}
with $\omega_\eta$ the volume element in  $\eta_{ab}$ and where it is
understood that  $\eta_{ab}$ and its inverse are used to raise and
lower tensor indices. 
It then follows that $\nabla_c ({}_n g_{ab})\equiv {}_n W_{cab} $ and
$\nabla_c g_{ab}\equiv W_{cab} $ are locally square
integrable. Therefore both ${}_n g_{ab}$ and $g_{ab}$ are regular metrics.

Now we can consider the limit $n\to \infty$.

{\bf Theorem:} Let ${}_n g_{ab}$ and  $ g_{ab}$ be regular metrics. Let
{\it i)}  ${}_n g_{ab}$ and  $ ({}_n g^{-1})^{ab}$ be locally uniformly
bounded and  {\it ii)} ${}_n g_{ab}$,  $ ({}_n g^{-1})^{ab}$ and the weak
derivative  $\nabla_c ({}_n g_{ab})$ converge locally in square integral to
$ g_{ab}$,  $ (g^{-1})^{ab}$ and  $\nabla_c  g_{ab}$
respectively. Then the corresponding curvature distributions ${}_n
R_{abc}^d $ converge to  $R_{abc}^d $ in the following sense: for any
test field $U^{abc}_d$,
\begin{equation}
\lim_{n\to\infty} {}_n R_{abc}^d[U^{abc}_d]=  R_{abc}^d[U^{abc}_d]
\end{equation}
(See \cite{gt87} for the proof).

It is straightforward to prove that (\ref{Anmetric}-\ref{Admetrici})
and (\ref{Anw},\ref{Aw}) satisfy the conditions of the above
theorem. We have

\begin{equation}
|({}_n g_{ab}| U^{ab})| \leq  |(S_{ab}| U^{ab})|
\end{equation}
and
\begin{equation}
|( ({}_n g^{-1})^{ab}| U_{ab})| \leq [2 \cosh(\beta x)]^2 |( (S^{-1})^{ab}| U_{ab})|
\end{equation}

It follows that ${}_n g_{ab}U^{ab}$ and $({}_n g^{-1})^{ab} U_{ab}$ are
bounded by smooth tensor fields with compact support, i.e. test
fields. Therefore ${}_n g_{ab}$ and $({}_n g^{-1})^{ab}$ are locally
uniformly bounded.

Let $U^{abcd}$ be a test tensor field on $R^4$. Define
\begin{equation}
\rho_n({}_ng,g)\equiv \int_{R^4}{({}_n g_{ab} - g_{ab})({}_n g_{cd} - g_{cd})
U^{abcd} \omega_\eta}
\end{equation}
It is easy to see that

\begin{equation}
\lim_{n\to\infty}\rho_n({}_ng,g)=0 \, .
\end{equation}
Then  ${}_n g_{ab}$  converges locally in square integral to
$g_{ab}$. The equivalent relation holds true for  $({}_n
g^{-1})^{ab}$. Finally,
let $U^{abcdef}$ be a test tensor field  on $R^4$. We have
\begin{equation}
\lim_{n\to\infty}
\int_{R^4}{({}_n W_{abc} - W_{abc})({}_n W_{cdf} - W_{cdf})
U^{abcdef} \omega_\eta} = 0 \, .
\end{equation}
Therefore  ${}_n W_{abc}$ converges locally in square integral to  $W_{abc}$.

\providecommand{\href}[2]{#2}\begingroup\raggedright\endgroup


\begin{thebibliography}{10}

\bibitem{rs99}
L.~Randall and R.~Sundrum,  {\em Phys.
  Rev. Lett.} {\bf 83} (1999) 4690--4693,
  \href{http://xxx.lanl.gov/abs/http://arXiv.org/abs/hep-th/9906064}{{\tt
  http://arXiv.org/abs/hep-th/9906064}}.

\bibitem{m98}
J.~Maldacena,  {\em Adv. Theor. Math. Phys.} {\bf 2} (1998) 231--252,
  \href{http://xxx.lanl.gov/abs/http://arXiv.org/abs/hep-th/9711200}{{\tt
  http://arXiv.org/abs/hep-th/9711200}}.

\bibitem{v83}
A.~Vilenkin,{\em Phys. Lett.}
  {\bf B133} (1983) 177--179.

\bibitem{is84}
J.~Ipser and P.~Sikivie,  {\em
  Phys. Rev.} {\bf D30} (1984) 712.

\bibitem{israel}
W.~Israel, 
  {\em Nuovo Cim.} {\bf B44S10} (1966) 1.

\bibitem{bcg99}
F.~Bonjour, C.~Charmousis, and R.~Gregory, 
  {\em Class. Quant. Grav.} {\bf 16} (1999) 2427--2445,
  \href{http://xxx.lanl.gov/abs/http://arXiv.org/abs/gr-qc/9902081}{{\tt
  http://arXiv.org/abs/gr-qc/9902081}}.

\bibitem{gt87}
R.~Geroch and J.~Traschen, {\em Phys. Rev.} {\bf D36} (1987) 1017.

\bibitem{g90}
G.~Goetz,
  {\em J.Math.Phys} {\bf 31} (1990) 2683. 

\bibitem{m93}
M.~Mukherjee,  {\em Class. Quant.
  Grav.} {\bf 10} (1993) 131.

\bibitem{gm99}
R.~Gass and M.~Mukherjee,  {\em
  Phys. Rev.} {\bf D60} (1999) 065011,
  \href{http://xxx.lanl.gov/abs/http://arXiv.org/abs/gr-qc/9903012}{{\tt
  http://arXiv.org/abs/gr-qc/9903012}}.

\bibitem{st99}
K.~Skenderis and P.~K. Townsend, {\em Phys. Lett.} {\bf B468} (1999) 46,
  \href{http://xxx.lanl.gov/abs/http://arXiv.org/abs/hep-th/9909070}{{\tt
  http://arXiv.org/abs/hep-th/9909070}}.

\bibitem{dfgk00}
O.~DeWolfe, D.~Z. Freedman, S.~S. Gubser, and A.~Karch,  {\em Phys. Rev.} {\bf D62} (2000)
  046008,
  \href{http://xxx.lanl.gov/abs/http://arXiv.org/abs/hep-th/9909134}{{\tt
  http://arXiv.org/abs/hep-th/9909134}}.

\bibitem{wl94} A.~Wong and P.~Letelier, {\em Phys.Rev.} {\bf D51}
(1995) 6612, {{\tt
  http://arXiv.org/abs/gr-qc/9411020}}.

\bibitem{cgs93}
M.~Cvetic, S.~Griffies, and H.~H. Soleng, {\em Phys. Rev.} {\bf D48} (1993)
  2613,
  \href{http://xxx.lanl.gov/abs/http://arXiv.org/abs/gr-qc/9306005}{{\tt
  http://arXiv.org/abs/gr-qc/9306005}}.

\bibitem{cgr92}
M.~Cvetic, S.~Griffies, and S.-J. Rey, {\em Nucl. Phys.} {\bf B381} (1992) 301,
  \href{http://xxx.lanl.gov/abs/http://arXiv.org/abs/hep-th/9201007}{{\tt
  http://arXiv.org/abs/hep-th/9201007}}.

\bibitem{b84}
W.~Boucher, {\em Nucl. Phys.} {\bf
  B242} (1984) 282.

\bibitem{t84}
P.~K. Townsend, {\em Phys. Lett.} {\bf B148} (1984)
  55.

\bibitem{bcy01}
F.~Brito, M.~Cvetic, and S.~Yoon, {\em Phys. Rev.} {\bf D64} (2001) 064021,
  \href{http://xxx.lanl.gov/abs/http://arXiv.org/abs/hep-ph/0105010}{{\tt
  http://arXiv.org/abs/hep-ph/0105010}}.

\bibitem{g99}
M.~Gremm, {\em Phys.
  Lett.} {\bf B478} (2000) 434--438,
  \href{http://xxx.lanl.gov/abs/http://arXiv.org/abs/hep-th/9912060}{{\tt
  http://arXiv.org/abs/hep-th/9912060}}.

\end{thebibliography}
\end{document}